\documentclass[twocolumn]{autart}

\usepackage{graphicx,psfrag}
\usepackage{amsmath,amssymb,amsfonts,mathrsfs}
\usepackage{color}
\usepackage{epsfig}
\usepackage{tabulary,enumerate}
\usepackage{algorithmic}
\usepackage{tkz-graph}

\newtheorem{assumption}{Assumption}
\newtheorem{assumption_d}{Assumption (Design)}
\newtheorem{definition}{Definition}
\newtheorem{theorem}{Theorem}
\newtheorem{lemma}{Lemma}

\newcommand{\R}{\mathbb{R}}

\newcommand{\eps}{\varepsilon}
\newcommand{\G}{\mathcal{G}}    %for graphs
\newcommand{\E}{\mathcal{E}}    %for graphs
\newcommand{\M}{\mathcal{M}}
\newcommand{\A}{\mathcal{A}}
\newcommand{\B}{\mathcal{B}}
\newcommand{\C}{\mathcal{C}}
\newcommand{\I}{\mathcal{I}}
\newcommand{\N}{\mathcal{N}}

\newcommand{\D}{\mathcal{D}}
\newcommand{\Q}{\mathcal{Q}}

\newcommand{\W}{\mathcal{W}}
\renewcommand{\S}{\mathcal{S}}    %for graphs

\renewcommand{\deg}{d}

\renewcommand{\S}{{ \mathcal{S} \stopmodif}}
\newcommand{\ave}{\operatorname{ave}} % 
\newcommand{\sign}{\operatorname{sign}} %

\newcommand{\dst}{\displaystyle}

\def\qedp{\hspace*{\fill}~{\tiny $\blacksquare$}}

\def\be{\begin{equation}}
\def\ee{\end{equation}}
\def\ba{\begin{array}}
\def\ea{\end{array}}
\def\eqa{\begin{eqnarray}}
\def\eqe{\end{eqnarray}}

\def\stopmodif{\color{black}}

\usepackage[prependcaption,colorinlistoftodos]{todonotes}

\begin{document}

\begin{frontmatter}

\title{Resilience against Misbehaving Nodes \\ in Asynchronous Networks}
% \thanksref{footnoteinfo}} 

%\thanks[footnoteinfo]{% This paper was not presented at any IFAC meeting. 
%Corresponding author P. Tesi E-mail: p.tesi@rug.nl} 

\author[Paestum]{D.M. Senejohnny}
\author[Rome]{S. Sundaram}   
\author[Paestum]{C. De Persis} 
\author[Rodi,Paestum]{P. Tesi} 

\address[Paestum]{
ENTEG, University of Groningen, 9747 AG Groningen, The Netherlands}      
\address[Rodi]{
DINFO, University of Florence, 50139 Florence, Italy}       
\address[Rome]{     
School of Electrical and Computer Engineering, Purdue University, West Lafayette, IN, USA}

\begin{abstract}
Network systems are one of the most active research areas in the engineering community as they feature a 
paradigm shift from centralized to distributed control and computation. 
When dealing with network systems, a fundamental challenge is to ensure their functioning 
even when some of the network nodes do not operate as intended due to faults or attacks.
The objective of this paper is to address the problem of resilient consensus in a context 
where the nodes have their own clocks, possibly operating in an asynchronous way, and can make updates at arbitrary time instants.
The results represent a first step towards the development of resilient event-triggered and self-triggered 
coordination protocols.
\end{abstract}
	
\end{frontmatter}	

%===============================================================================

\section{Introduction}

Network systems are one of the most active research areas
in the engineering community as they feature a paradigm shift 
from centralized to distributed control and computation.  When dealing with 
network systems, a fundamental challenge is to ensure 
their functioning even when some of the network units (nodes)
do not operate as intended due to faults or attacks.
The main difficulty originates from the fact that  
normal (non-misbehaving) nodes can receive, process, and spread
erroneous data coming from misbehaving nodes with the consequence that
a failure in one point of the network can compromise the whole network functioning. 

The prototypical problem to study resilience in the presence of misbehaving nodes
is the so-called consensus problem \cite{cao2013overview}, 
which forms the foundation for distributed computing.
In \emph{resilient consensus}, each node is assumed to be aware of only local information 
available from its neighbors and the goal is to make sure that 
normal nodes eventually reach a common value despite 
the presence of misbehaving nodes. The resilient consensus 
problem has a long history, and it has been investigated first by computer scientists 
\cite{Dolev1986,Lynch}, usually under the hypothesis 
that the network graph is complete, that is assuming an all-to-all communication structure.
More recently, thanks to the widespread of consensus-based applications, this problem
has attracted a lot of interest also within the engineering community, 
mostly in connection with the goal of delineating the minimal connectivity hypotheses  
that are needed to secure consensus.

In \cite{Sundaram2013}, the authors consider \emph{mean subsequence reduced} (MSR) algorithms
and define a graph-theoretic property, referred to as \emph{network robustness},
which charactherizes necessary and sufficient connectivity hypotheses under which
normal nodes can reach consensus using only local information available from their neighbors.
The results indicate that, while the communication graph should possess a certain degree of redundancy,
completeness of the communication graph is not necessary even for very general types of misbehavior.
The results of \cite{Sundaram2013} have been extended in many venues. Examples include
methods for handling time-varying networks \cite{SaldanaACC2017}, 
double-integrator systems \cite{DibajiSCL2015},
sparse communication graphs \cite{Abbas2014} as well 
as methods for identifying the robustness of specific classes of networks \cite{Usevitch2017}.
 
Most of the research works in this area assume that the network operates
in perfect \emph{synchrony}, in the sense that all the nodes, at least the normal ones, 
update at the same moment in time. Since this condition might be difficult to obtain,
a parallel line of research has focused on methods for handling \emph{asynchrony}, 
which is known to render consensus much more challenging 
to obtain \cite{Dolev1986}. Among many notable works,
we mention \cite{LeBlanc2012,Vaidya2012,DibajiAUT2017,DibajiTAC2017}
which consider MSR-type algorithms supporting asynchrony.
In these works, asynchrony refers to the property that the nodes are equipped with 
identical clocks, operating synchronously, 
but can make updates at different {steps},
that is at different multiples of the clock period. 

The objective of this paper is to address the problem of resilient consensus 
in a context where the nodes have their own clocks, possibly operating in 
an asynchronous way, and can make updates at arbitrary time instants.
Besides the practical difficulties in achieving a perfect clock synchronization, 
one main reason for considering independent clocks is related to
developments in the area of networked control systems where, in order to enhance
efficiency and flexibility, it is more and more required to have fully autonomous devices, which 
is the paradigm of
\emph{event-triggered} and \emph{self-triggered} control \cite{heemels2012introduction}.
In fact, our approach utilizes a self-triggered control scheme \cite{de2013robust}.
Each node is equipped with a clock that determines when the next update is
scheduled. At the update instant, the node polls its neighbors, collects the data
and determines whether it is necessary to modify its controls along with a bound
on the next update instant.

The main result of this paper establishes \emph{approximate} consensus
under certain conditions on the connectivity of the communication graph
and a maximum number of misbehaving nodes (Theorem \ref{thm_1}), conditions
which can be relaxed if misbehavior only occurs in data acquisition 
or timing (Theorem \ref{thm_2}). While  \cite{Sundaram2013,LeBlanc2012,DibajiAUT2017,DibajiTAC2017}
achieve perfect consensus and require milder connectivity conditions,
the present results indicate that the resilient consensus problem can be approached 
without requiring that the nodes are equipped with identical clocks, 
even when the graph is not complete, a feature which is very appealing 
for networked control applications.

From a technical viewpoint, we adopt a control logic which 
removes ``extreme" values (as in classic MSR-type algorithms) and then form an average
from a subset of the remaining values through 
a \emph{quantized sign} function, which saturates the control action
applied at the node. This is as an approximation of the 
pure (non-quantized) control law introduced in \cite{cortes2006finite},
which, in the absence of misbehaving nodes, guarantees  
\emph{max-min} consensus, the quantization being instrumental 
to avoid a continuous data flow among the nodes. Interestingly, the use of 
sign functions has been considered to solve consensus on the \emph{median value}
\cite{FranceschelliTAC2017}, which is inherently robust to outliers
and thus to some types of misbehavior. Although our work
is substantially different from \cite{FranceschelliTAC2017} as we do not 
consider a continuous data flow, both the approaches
suggest that saturating the controls can be beneficial for resilience
since this limits the effect of an incorrect update choice
resulting from erroneous data. 
 
\section{System definition and main result}

Consider a network of $n \in \mathbb N$ nodes interconnected  
in accordance with a time-invariant undirected connected 
graph $\G:=(\I,\E)$, where $\I$ is the 
set of nodes while $\E \subseteq \I \times \I$ is the set of edges. 
We let $\Q_i$ denote the set of neighbors of $i \in \I$,
and by $\deg_i$ the cardinality of $\Q_i$, that is $\deg_i := |\Q_i|$.
The set $\Q_i$ represents the set of nodes with which node $i$ exchanges data.
For every $i \in \I$, the dynamics are given by
\begin{eqnarray} \label{eq:sys}
\left\{
\def\arraystretch{1.5}
\begin{array}{l}
\dot x_i(t) = u_i(t) \\  
z_i(t) = f_i(x_i(t))
\end{array}
\right.
\quad t \in \R_{\geq 0} 
\end{eqnarray}
where $x_i \in \R$ is the state with $x_i(0)$ arbitrary; $u_i \in \R$ is the control action applied by 
node $i$; $z_i \in \R$ is the output, where $f_i: \R \rightarrow \R$
is a function to be specified, and 
represents the value that node $i$ makes available to its neighbors.
The variable $t \in \R_{\geq 0}$  is understood as the \emph{absolute} time frame
within which all the nodes carry out their operations in an asynchronous way.

The objective is to design a coordination protocol in such a way that 
{normal} (non-misbehaving) nodes eventually reach \emph{approximate} consensus
despite the presence of misbehaving nodes. We will specify later on the class of misbehaviors 
considered in this paper. According to the usual notion of consensus \cite{cao2013overview}, 
the network nodes should converge  
to an equilibrium point where all the nodes have the 
same value lying somewhere between the minimum and maximum of their initial values.  
The following definition formalizes the notion of \emph{approximate} consensus considered in this paper.

\begin{definition}\label{def:approx_consensus}
The network is said to reach approximate consensus 
if, for every initial value of the nodes, 
normal nodes remain between the minimum and maximum of their initial values,
and there exists a constant $c \in \mathbb R_{> 0}$ such that
$\limsup_{t \rightarrow \infty} |x_i(t)-x_j(t)| < c$ for every pair $(i,j)$
of {normal} nodes.
\qedp
\end{definition}

Network nodes carry out their operations by means of three main quantities:
\begin{itemize}
\item A parameter $\eps \in \mathbb R_{> 0}$, which determines the desired
level of accuracy for consensus.
\item A parameter $F \in \mathbb N$, which determines the maximum number  
of misbehaving nodes that the network is expected to encounter.
\item A sequence $\{t^i_k\}_{k \in \mathbb N}$ of time instants at which
node $i$ requests data from its neighbors, where $t^i_0 \in [0,t_{init}]$ defines
the first time instant at which node $i$ becomes active and $t_{init} \in \mathbb R_{\geq 0}$
denotes the first time instant at which
all the nodes are active in the network. 
By convention, $0=t^r_0$ where $r$ is the first network node to become active
and $x_i(t)=x_i(t^i_0)$ for every $i \in I$ and for all $t \in [0,t^i_0]$.
\end{itemize}

It is implicit in the above definition of $t_{init}$ that \emph{all} the nodes 
become active in a finite time. We will also assume that all nodes remain active for the entire runtime.
The analysis can be easily generalized to the case where some of the nodes 
never ``wake up" or ``die" during the network runtime.

\subsection{Coordination protocol}

Let $\N$  and $\M$ represent the sets of normal nodes and misbehaving nodes,
respectively, which are assumed to be time-invariant (Assumption 1).
We now focus on the generic $k$-th round of operations for node $i \in \N$.
This consists of four main operations: \emph{(i) data acquisition};
\emph{(ii) data transmission}; \emph{(iii) control logic}; \emph{(iv) timing}. 
These operations will also define the considered notion of misbehaviors.

\emph{(i) Data acquisition.} 
At time $t^i_k$, node $i \in \I$ collects data from its neighbors. 
Denote by $h_i: \R \rightarrow \R$, $i \in \I$, the function processing the
incoming data, which means that given $z_j(t)$ with $j \in \Q_i$, 
$h_i(z_j(t))$ defines the information on $j$ available to node $i$ at time $t$.
For $i \in \N$, 
\begin{eqnarray} \label{eq:sensing}
h_i(\chi) = \chi \quad \forall \chi \in \R
\end{eqnarray}
A data acquisition error means that (\ref{eq:sensing}) is not satisfied 
for some $\chi \in \R$, which represents for
example a fault at the receiver. 

\emph{(ii) Data transmission.} 
For $i \in \N$, $f_i$ in (\ref{eq:sys}) satisfies 
\begin{eqnarray} \label{eq:transmission}
f_i(\chi) = \chi \quad \forall \chi \in \R
\end{eqnarray}
which means that a normal node makes available to the other nodes
its true state value. A transmission error means that (\ref{eq:transmission}) is not satisfied 
for some $\chi \in \R$, which can represent 
a fault at the transmitter as well as an intentional
misbehavior. By convention, node $i$ transmits data from time $t^i_0$.

\emph{(iii) Control logic.} 
The scheme is based on the idea of discarding ``extreme" values \cite{Dolev1986},
which prevents normal nodes from processing potentially harmful information. 
For every $i \in \I$, let $\D_i(t) \subseteq \Q_i$ 
be the set of neighbors that are not discarded by $i$ at $t \in \R_{\geq 0}$.
For $i \in \N$ this set is determined as follows.
Let ${\mathcal V}_i(t)$ be the ordered set formed 
by sorting the elements of $\Q_i$ in a non-decreasing order of value $h_i(z_j(t))=z_j(t)$.
An arbitrary ordering is pre-specified to classify elements with the same value.
Consider the set ${\mathcal F}_i(t)$ 
formed by the first $F$ elements of ${\mathcal V}_i(t)$, and let 
$\underline {\mathcal E}_i(t)$ be the subset of ${\mathcal F}_i(t)$ consisting 
of all the elements of ${\mathcal F}_i(t)$ with associated value smaller than $x_i(t)$,
that is $r \in \underline {\mathcal E}_i(t)$ if and only if $r \in {\mathcal F}_i(t)$
and $z_r(t) < x_i(t)$. Similarly, let ${\mathcal L}_i(t)$ be the set
formed by the last $F$ elements of ${\mathcal V}_i(t)$, and let 
$\overline {\mathcal E}_i(t)$ be the subset of ${\mathcal L}_i(t)$ consisting 
of all the elements of ${\mathcal L}_i(t)$ with associated value larger than $x_i(t)$,
that is $r \in \overline {\mathcal E}_i(t)$ if and only if $r \in {\mathcal L}_i(t)$
and $z_r(t) > x_i(t)$. By convention, $r \notin \D_i(t)$ if $t<t^r_0$, that is if node $r$ is
still not active at time $t$. For $i \in \N$, $\D_i(\cdot)$ satisfies
\begin{eqnarray} \label{eq:D}
\D_i(t) = \Q_i \, \backslash \, \left( \underline {\mathcal E}_i(t) \cup \overline {\mathcal E}_i(t) \right)
\end{eqnarray}
and the control action is given by
\begin{eqnarray} \label{eq:controls}
u_i(t) =
\left\{
\begin{array}{ll}
0 & \qquad t \in [0,t^i_0) \\ 
\displaystyle \sign_\eps(\ave_i(t^i_k)) & \qquad
t \in [t^i_k,t^i_{k+1})
\end{array} \right.
\end{eqnarray}
where
\begin{eqnarray} \label{eq:ave}
\ave_i(t) := \sum_{j \in \D_i(t)} \left( h_i(z_j(t))-x_i(t) \right) 
\end{eqnarray}
and where, for every $\chi \in \R$,
\begin{eqnarray} \label{eq:signeps}
\sign_\eps(\chi) :=
\left\{
\begin{array}{ll}
0 & \qquad \textrm{if } |\chi| < \eps \\ 
\sign(\chi) &  \qquad \textrm{otherwise}
\end{array} \right.
\end{eqnarray}
By convention, $\D_i(t)=\emptyset$ implies $\ave_i(t)=0$. 
An error in the control logic means that (\ref{eq:controls}) is not satisfied 
for some $t \in \R_{\geq 0}$.

\emph{(iv) Timing.} 
For $i \in \N$, the next round of operations is scheduled at time
$t^i_{k+1} = t^i_k + \Delta_k^i$, where
{\setlength\arraycolsep{2pt}
\def\arraystretch{1}
\begin{eqnarray} \label{eq:clock} 
\begin{array}{l}
\Delta_k^i \geq \underline \Delta_i \\ \\ 
\Delta_k^i  \leq \dst\frac{1}{4 d_i}\, \max \{ \eps, \, |\ave_i(t^i_k)| \}
\end{array}
\end{eqnarray}}%
with $\underline \Delta_i \in \R_{>0}$ such that $\underline \Delta_i \leq \eps/(4d_i)$.
Operations can be then periodic as well as aperiodic. The first condition 
avoids arbitrarily fast sampling (Zeno behavior), while the second of condition is needed
to reach approximate consensus. A timing error means that (\ref{eq:clock}) is not satisfied 
for some $k \in \mathbb N$.

\subsection{Assumptions and main results} \label{sec:ass}

\begin{assumption} \label{ass:M}
The set $\M$ of misbehaving nodes does not change over time and 
$|\M| \leq F$. \qedp
\end{assumption}

\begin{assumption} \label{ass:solutions}
For every $i \in \M$, $u_i(\cdot)$ is a locally integrable function,
$\D_i(\cdot) \subseteq \Q_i$, $f_i(\cdot),h_i(\cdot) \in \R$ and
$\Delta_k^i \geq \underline \Delta_i$ for all $k \in \mathbb N$
for some $\underline \Delta_i \in \R_{>0}$.
\qedp
\end{assumption}

\setcounter{assumption_d}{2}
\begin{assumption_d} \label{ass:connect_1}
Every pair of normal nodes have at least $3 F + 1$ neighbors in common.
\qedp
\end{assumption_d}

\begin{assumption_d} \label{ass:connect_2}
Every pair of normal nodes have at least $2 F + 1$ neighbors in common.
\qedp
\end{assumption_d}

The second assumption ensures the existence of the solutions for all the nodes {and for all time,} that
variables and functions are well defined. 
Assumption \ref{ass:solutions} entails 
no upper bound on $\Delta^i_k$. This is {in} order to capture the event that a misbehaving node never 
collects data from its neighbors and applies an open-loop control.

Assumptions \ref{ass:connect_1} and \ref{ass:connect_2} 
deal with the graph connectivity properties, and their use will vary 
depending on the type of nodes misbehavior.
Both the assumptions ensure that the normal nodes share sufficient 
``genuine" information for taking control decisions. These assumptions hold, 
for instance, for classes of \emph{strongly regular graphs} \cite{Godsil},
though it is not needed that the graph is {regular}. 
These assumptions should be interpreted as design conditions
when the graph topology can be assigned.
{
\color{black}
The connectivity conditions in Assumptions  \ref{ass:connect_1} and \ref{ass:connect_2} are not difficult to check and, for large values of $n$, can also lead to sparse network configurations.
 Let $\lambda$ denote the number of neighbors that every pair of normal node share. From the Assumptions \ref{ass:connect_1} and \ref{ass:connect_2}, $\lambda$ is either $3F+1$ or $2F+1$. Consider a complete graph of $\lambda+1$ nodes.  Then add $k$ more nodes, where each of these $k$ nodes connects to all of the $\lambda+1$ nodes in  the clique.  This graph with $n=\lambda+k+1$ nodes has the property that every pair of nodes has at least $\lambda$ nodes in common and we see that, for fixed $F$, the number of edges scales only linearly with $n$. 
%This indicates that the connectivity conditions in Assumptions  \ref{ass:connect_1} and \ref{ass:connect_2} are not difficult to check and, for large values of $n$, can also lead to sparse network configurations.
}

We now state the main results of the paper, which are proven in Sections 4 and 5.

\begin{theorem} \label{thm_1}
Consider the network system (\ref{eq:sys})-(\ref{eq:clock}), with the misbehaving nodes exhibiting 
an error in any of the operations (i)-(iv). 
If Assumptions \ref{ass:M}, \ref{ass:solutions} and \ref{ass:connect_1} hold true, 
then all the normal nodes $i\in \mathcal{N}$ remain inside the convex hull 
containing their initial values. Moreover, there exists 
a finite time $T \in \R_{\geq 0}$ 
such that $|x_i(t) - x_j(t)| < 3\eps$ for all $t \geq T$ and $i,j \in \N$. \qedp 
\end{theorem}

\begin{theorem} \label{thm_2}
Consider the network system (\ref{eq:sys})-(\ref{eq:clock}), with the misbehaving nodes exhibiting 
an error in the operation (i) and/or (iv). 
If Assumptions \ref{ass:M}, \ref{ass:solutions} and \ref{ass:connect_2} hold true, 
then all the normal nodes $i\in \mathcal{N}$ remain inside the convex hull 
containing their initial values. Moreover, there exists 
a finite time $T \in \R_{\geq 0}$ 
such that $|x_i(t) - x_j(t)| < 3\eps$ for all $t \geq T$ and $i,j \in \N$. \qedp 
\end{theorem}

Intuitively, errors in data acquisition or timing are less critical
as they do not alter control or output values. 

%%%%%%%%%%%%%%%%%%%%%%%%%%%%%%%%%%%%%%%%%%%%%%%%%%%%
%%%%%%%%%%%%%%%%%%%%%%%%%%%%%%%%%%%%%%%%%%%%%%%%%%%%

\section{Monotonicity properties}

The results of this section
rely on Assumption \ref{ass:M} and \ref{ass:solutions} only, and are thus 
independent of the specific type of nodes misbehavior.
Let
\begin{eqnarray}
x_m(t) := \min_{i \in \N} x_i(t), \quad 
x_M(t) := \max_{i \in \N} x_i(t)
\end{eqnarray} 
where $t \in \R_{\geq 0}$. 

The first result shows that normal nodes remain 
in the convex hull containing their initial values. 

\begin{lemma} \label{lem:tec_1}
Consider the network system (\ref{eq:sys})-(\ref{eq:clock}), and let 
Assumptions \ref{ass:M} and \ref{ass:solutions}
hold. Then, $x_m(\cdot)$ and $x_M(\cdot)$
are monotonically non-decreasing and non-increasing, respectively.
\end{lemma}

\emph{Proof}.
We prove the statement only for $x_m(\cdot)$ since the proof for $x_M(\cdot)$ 
is analogous. Suppose that the claim is false, and let $\tau$ be the first time instant
at which there exists an index $i \in \N$ such that 
{\setlength\arraycolsep{2pt}
\begin{eqnarray} \label{lem:tec_1:eq_1}
\left\{
\def\arraystretch{1.5}
\begin{array}{l}
x_i(\tau) \leq x_j(\tau) \quad \forall \, j \in \N \\
u_i(\tau) < 0
\end{array}
\right.
\end{eqnarray}}%
Clearly, there could be multiple nodes achieving (\ref{lem:tec_1:eq_1}) at time $\tau$.
In this case, $i$ is any of such nodes. Notice that $\tau \geq t^i_0$ since 
$u_i(t)=0$ for all $t \in [0,t^i_0)$.

Consider first the case where $\tau = t^i_k$ for some $k \in \mathbb N$.
In order for $u_i(t^i_k) < 0$ we must have $\ave_i(t^i_k) \leq - \eps < 0$.
However, this is not possible. In fact, any normal node $j$ satisfies 
$z_j(t^i_k) = x_j(t^i_k) \geq x_i(t^i_k)$ because $i$ is the node of minimum value at $\tau = t^i_k$. 
Hence, $z_j(t^i_k) < x_i(t^i_k)$ only if $j$ is misbehaving.
Since misbehaving nodes are not more than $F$ by Assumption \ref{ass:M},
if a misbehaving node $j$ gives $z_j(t^i_k) < x_i(t^i_k)$
it is discarded by the control logic. 

Consider next the case where $\tau$ is not an update time for node $i$.
Let $t^i_k < \tau$ be the last update time for node $i$ before $\tau$.
In order to have (\ref{lem:tec_1:eq_1}), there must exist
a node $s \in \N$ such that 
{\setlength\arraycolsep{2pt}
\begin{eqnarray} \label{lem:tec_1:eq_2}
\left\{
\def\arraystretch{1.5}
\begin{array}{l}
x_s(t^i_k) \leq x_j(t^i_k) \quad \forall \, j \in \N \\
x_s(t^i_k) < x_i(t^i_k)  \\
x_i(\tau) \leq x_j(\tau) \quad \forall \, j \in \N \\
u_i(\tau) < 0 
\end{array}
\right.
\end{eqnarray}}%
The first two conditions imply that $i$ is not the node which takes on the minimum
value at $t^i_k$, value which is instead attained by node $s$. Condition $x_s(t^i_k) < x_i(t^i_k)$ is needed
otherwise $u_i(t^i_k) \geq 0$ in accordance with 
the previous arguments. 
The last three 
conditions mean that $i$ becomes the minimum at $\tau$ with $u_i(\tau) = u_i(t^i_k) < 0$. 
Let $\beta := x_i(t^i_k) - x_s(t^i_k)$.
Recall that normal nodes take controls in $\{-1,0,1\}$. Hence,
$x_i(\tau) \leq x_j(\tau)$ for all $j \in \N$ only if
$\tau-t^i_k \geq \beta/2$.
However, 
{\setlength\arraycolsep{2pt}
\begin{eqnarray} \label{lem:tec_1:eq_3}
\Delta^i_k 
&\leq& \dst\frac{1}{4 d_i} \sum_{j \in \D_i(t^i_k)} (x_i(t^i_k)-z_j(t^i_k)) \nonumber \\
&\leq& \dst\frac{1}{4 d_i} \sum_{j \in \D_i(t^i_k)} (x_i(t^i_k)-x_s(t^i_k)) 
\end{eqnarray}}%
The first equality follows from the fact that $u_i(t^i_k) < 0$ requires
$\ave_i(t_k^i) \leq -\eps$ so that $\Delta^i_k \leq |\ave_i(t^i_k)|/(4d_i)$.
On the other hand, the second inequality follows since
$z_j(t^i_k) < x_s(t^i_k)$ only if $j$ is misbehaving,
in which case it is
discarded by node $i$ in view of Assumption \ref{ass:M} and by the control logic.
Since $|\D_i(t)| \leq d_i$ for all $t \in \R_{\geq0}$ then $\Delta^i_k \leq \beta/4$. This leads to 
a contradiction since it implies $\beta/2 \leq \tau-t^i_k< \Delta^i_k \leq \beta/4$
with $\beta>0$. \qedp

By Lemma \ref{lem:tec_1}, normal nodes remain in the convex 
hull containing their initial values. 
This lemma also implies that $x_m(\cdot)$ and $x_M(\cdot)$ admit a finite limit,
{\setlength\arraycolsep{2pt}
\begin{eqnarray} 
\underline x := \lim_{t \rightarrow \infty} x_m(t), \quad
\overline x := \lim_{t \rightarrow \infty} x_M(t)
\end{eqnarray}}%
For the next developments, we strengthen Lemma \ref{lem:tec_1} by showing  
that there exist normal nodes that settle on the minimum and maximum values 
in a finite time.

\begin{lemma} \label{lem:tec_2}
Consider the network system (\ref{eq:sys})-(\ref{eq:clock}), and let 
Assumptions \ref{ass:M} and \ref{ass:solutions}
hold. Then, there exist at least two indices $r,s \in \N$ 
and a finite time $T' \in \R_{\geq 0}$ such that $x_r(t)=\underline x$ and $x_s(t)=\overline x$
for all $t \geq T'$. In addition, $\min_{i \in \N} x_i(t) \geq \underline x$ and 
$\max_{i \in \N} x_i(t) \leq \overline x$ for all $t \geq T'$. 
\end{lemma}

\emph{Proof}. 
We prove the statement only for $\underline x$ as the proof for $\overline x$ 
is analogous. Since $x_m(\cdot)$ converges to $\underline x$ and is continuous,
for any $\delta \in \R_{>0}$ there exists a finite time $T_\delta \in \R_{\geq 0}$ such that 
$|x_m(t) - \underline x| < \delta$ for all $t > T_\delta$. Let $\underline \Delta := \min_{i \in \N} \underline \Delta_i$
and pick $\delta = \underline \Delta/3$.
Consider any $i \in \N$ and any update time $t^i_k$ for node $i$ such that $t^i_k \geq T_\delta$. 
Condition $t^i_k \geq T_\delta$ is well defined 
for any $T_\delta$ since by Lemma \ref{lem:tec_1}
normal nodes always remain in the convex 
hull containing their initial values so that $\Delta^i_k$ is bounded from above.

We claim that, for any $i \in \N$ and any $t^i_k \geq T_\delta$, 
{\setlength\arraycolsep{2pt}
\begin{eqnarray} \label{lem:tec_2:eq_1}
|x_i(t^i_k) - \underline x | \geq \delta \,\, \Rightarrow \,\, 
|x_i(t) - \underline x | \geq \delta \quad \forall \, t \geq t^i_k
\end{eqnarray}}%
In simple terms, this means that if $x_i(t^i_k)$ does not belong to 
$W:=(\underline x - \delta,\underline x + \delta)$ then $x_i(\cdot)$
can never enter $W$ afterwards. 
The implication \eqref{lem:tec_2:eq_1}
is shown as follows. Since $|x_m(t) - \underline x| < \delta$ for all $t > T_\delta$ then we must also have
$x_j(t)  > \underline x - \delta$ for all $t > T_\delta$ and $j \in \N$. This means that
condition $|x_i(t^i_k) - \underline x | \geq \delta$ implies $x_i(t^i_k) \geq \underline x + \delta$.
The analysis is divided into two subcases.

\emph{Case 1}. Assume 
$x_i(t^i_k) \in [\underline x + \delta, \underline x + 2 \delta)$.
In this case, 
{\setlength\arraycolsep{2pt}
\begin{eqnarray} \label{lem:tec_2:eq_2}
\ave_i(t^i_k) &=& \sum_{j \in \D_i(t^i_k)} (z_j(t^i_k)-x_i(t^i_k)) \nonumber \\
&>& \sum_{j \in \D_i(t^i_k)} (\underline x -\delta - x_i(t^i_k)) \nonumber \\
&>& -3\delta |\D_i(t^i_k)|  \nonumber \\
&>& - \eps 
\end{eqnarray}}%
The first inequality follows from the fact that $j \in \D_i(t^i_k)$ only if $z_j(t^i_k) > \underline x - \delta$.
In fact, $z_j(t) = x_j(t)  > \underline x - \delta$ for all $t > T_\delta$ and $j \in \N$.
Thus, nodes with an output value less than or equal to $\underline x - \delta$ are misbehaving,
and they are discarded in view of Assumption \ref{ass:M} and by 
construction of the control logic. The second inequality follows because 
$x_i(t^i_k) < \underline x + 2 \delta$ by hypothesis. The last inequality follows since
$3 \delta \leq \eps/(4d_i)$ and $|\D_i(t)| \leq d_i$ for all $t \in \R_{\geq0}$.
Since $\ave_i(t^i_k)>-\eps$ implies $u_i(t) \in \{0,1\}$ for all $t \in T^i_k$ then
$x_i(t) \notin W$ for all $t \in T^i_k$,
and $x_i(t^i_{k+1}) \notin W$ by continuity of $x_i(\cdot)$.

\emph{Case 2}. Assume $x_i(t^i_k) \geq \underline x + 2 \delta$. In order for node $i$
to decrease we must have $\ave_i(t^i_k) \leq -\eps$. Hence, 
{\setlength\arraycolsep{2pt}
\begin{eqnarray} \label{lem:tec_2:eq_3}
\Delta^i_k 
&\leq& \dst\frac{1}{4 d_i} \sum_{j \in \D_i(t^i_k)} (x_i(t^i_k)-z_j(t^i_k)) \nonumber \\
&<& \dst\frac{1}{4 d_i} \sum_{j \in \D_i(t^i_k)} (x_i(t^i_k)- \underline x + \delta) \nonumber \\
&\leq& \dst\frac{1}{4} (x_i(t^i_k)- \underline x + \delta)
\end{eqnarray}}%
The second inequality follows because 
$j \in \D_i(t^i_k)$ only if $z_j(t^i_k) > \underline x - \delta$ according with the previous arguments.
The third inequality follows since 
$|\D_i(t)| \leq d_i$ for all $t \in \R_{\geq0}$. 
Since normal nodes take controls in $\{-1,0,1\}$, we obtain
{\setlength\arraycolsep{2pt}
\begin{eqnarray} \label{lem:tec_2:eq_4}
x_i(t) &\geq& x_i(t^i_k) - \Delta^i_k \geq  \underline x + \frac{5}{4} \delta 
\end{eqnarray}}%
for all $t \in T^i_k$, where the last inequality follows from (\ref{lem:tec_2:eq_3}) 
and since $x_i(t^i_k) \geq \underline x + 2 \delta$. Thus $x_i(t) \notin W$ for all $t \in T^i_k$,
and $x_i(t^i_{k+1}) \notin W$ by continuity of $x_i(\cdot)$.

We conclude that if $x_i(t^i_k)$ does not belong to 
$W$ for some $t^i_k \geq T_\delta$ then $x_i(\cdot)$
can never enter $W$ afterwards. Moreover, for every $i \in \N$, 
if $x_i(t^i_k) \in W$ and $u_i(t^i_k) \neq 0$ then
$x_i(t^i_{k+1}) \notin W$. In fact, in this case,
node $i$ must apply the same control input for a period not shorter than $\underline \Delta$. Thus,
$x_i(t^i_{k+1}) \notin W$ since the control input is constant 
with unitary slope for at least $\underline \Delta$ time units
and $W$ has measure $2\delta = 2\underline \Delta/3$. 
Thus, since the number of nodes is finite, 
there exists a finite time $T'' \geq T_\delta$ starting from which the signal $x_i(\cdot)$,
$i \in \N$, either persistently remains inside $W$ or persistently
remains outside $W$. Moreover, there exists at least 
one index $i \in \N$ for which $x_i(\cdot)$ persistently remains inside $W$
since, by definition, $\underline x \in W$ is the limiting value of $x_m(\cdot)$.

Every $x_i(\cdot)$, $i \in \N$, 
that persistently remains outside $W$ from $T''$ onwards satisfies
$x_i(t) \geq \underline x +\delta$ for all $t \geq T''$. 
Consider next any $x_i(\cdot)$, $i \in \N$,  that persistently remains inside $W$
from $T''$ onwards.
By the above arguments, $u_i(t^i_k) = 0$ for all $t^i_k \geq T''$. 
Moreover, the first sampling $t^i_k \geq T''$ must occur no later than 
$T' := T'' + \eps/4$.
This is because, either $u_i(T'') = 0$ so that $t^i_k - T'' \leq \eps/(4d_{i})$ 
by construction of the update times
or $u_i(T'') \neq 0$ so that $t^i_k - T'' \leq 2\delta < \eps/4$
otherwise $x_i(t^i_{k}) \notin W$ according to the previous arguments. 
Hence, every $x_i(\cdot)$, $i \in \N$, that persistently remains inside $W$ from $T''$ onwards
satisfies $x_i(t) = x_i(T')$ for all $t \geq T'$. Thus
$\min_{i \in \N} x_i(t) \geq \underline x$ for all $t \geq T'$ and 
$x_r(T')=\underline x$ for some $r \in \N$ since
$\underline x$ is the limiting value of $x_m(\cdot)$. 
\qedp

%%%%%%%%%%%%%%%%%%%%%%%%%%%%%%%%%%%%%%%%%%%%%%%%%%%%%%%
%%%%%%%%%%%%%%%%%%%%%%%%%%%%%%%%%%%%%%%%%%%%%%%%%%%%%%%
\section{Generic misbehavior}

By Lemma \ref{lem:tec_2}, there exist at least two indices $r,s \in \N$ 
and a finite time $T'$ such that $x_r(t)=\underline x$ and $x_s(t)=\overline x$
for all $t \geq T'$. 
We now show that under Assumption \ref{ass:connect_1} $\overline x- \underline x$ is upper bounded 
by $3 \eps$. 

\emph{Proof of Theorem 1}. The property that normal nodes always remain 
inside the convex hull containing their initial values has been shown in Lemma \ref{lem:tec_1}.
Thus, we focus on the second part of the statement.  

Let $T'$ be as in Lemma \ref{lem:tec_2},
and denote by $r$ and $s$ any two indices belonging to $\N$ such that 
$x_r(t)=\underline x$ and $x_s(t)=\overline x$ for all $t \geq T'$.
Consider now any update time $t^r_k \geq T'$ for node $r$.
Since node $r$ does not change its value from $T'$ onwards, we must have
$\ave_r(t^r_k) < \eps$. This implies that $z_i(t^r_k)-x_r(t^r_k)<\eps$
for all $i \in \D_r(t^r_k)$. 
In fact, in order to have $z_i(t^r_k)-x_r(t^r_k) \geq \eps$
for some $i \in \D_r(t^r_k)$ there should exist at least one index $j \in \D_r(t^r_k)$
such that $z_j(t^r_k)-x_r(t^r_k) < 0$.
However, this is not possible.
In fact, in view of Lemma \ref{lem:tec_2}, $z_j(t^r_k)=x_j(t^r_k) \geq \underline x$ for all $j \in \N$
so that every node which takes on
an output value smaller than $\underline x$ is misbehaving and it is discarded by node $r$ in view
of Assumption \ref{ass:M} and by construction of the control logic.
Since every $i \in \N$ satisfies $z_i(\cdot) \equiv x_i(\cdot)$ 
we have $x_i(t^r_k)-\underline x<\eps$ for all $i \in \D_r(t^r_k) \cap \N$.

The claim thus trivially follows if $s \in \D_r(t^r_k)$. Suppose
$s \notin \D_r(t^r_k)$. Nodes $r$ and $s$ have  at least $3 F +1$
neighbors in common by virtue of Assumption \ref{ass:connect_1}, 
so that, because of Assumption \ref{ass:M}, they have at least $2 F+1$ neighbors in common 
belonging to $\N$. At $t^r_k$, node $r$ can discard at most 
$F$ normal nodes since there are no normal nodes with value smaller than
$x_r(t^r_k)$. Hence, there are at least $F+1$
normal nodes belonging to $\D_r(t^r_k) \cap \Q_s$. 
Starting from $t^r_k$,
node $s$ will sample at a time $t^s_k \leq t^r_k + \eps/4$.
When $s$ will sample,
it cannot discard all these $F+1$ normal nodes
by construction of the control logic 
and because none of these nodes can take on a value larger than $\overline x$ from 
$T'$ onwards in view of Lemma \ref{lem:tec_2}. Following the same reasoning as before,
at least one of these nodes, say node $i$, must satisfy
$x_i(t^s_k) > \overline x - \eps$  otherwise one would have $\ave_s(t^s_k) \leq -\eps$. 
Hence, 
{\setlength\arraycolsep{2pt}
\begin{eqnarray} \label{thm_1:eq_1}
\left\{
\def\arraystretch{1.5}
\begin{array}{l}
x_i(t^r_k)  < \underline x +  \eps \\
x_i(t^s_k) > \overline x - \eps  \\
x_i(t^s_k) \leq x_i(t^r_k) + \dst \frac{\eps}{4}
\end{array}
\right.
\end{eqnarray}}%
where the last inequality follows  since $u_i(t)<1$
for all $t \in \R_{\geq 0}$ and $i \in \N$, and since $t^s_k \leq t^s_k+\eps/4$.
This implies $\overline x - \underline x < 3 \eps$, and 
the claim follows letting $T:=T'$. \qedp
 
%%%%%%%%%%%%%%%%%%%%%%%%%%%%%%%%%%%%%%%%%%%%%%%%%%%%%%%%%%%%%%%%%%%%%%%%%%%%%%%%
%%%%%%%%%%%%%%%%%%%%%%%%%%%%%%%%%%%%%%%%%%%%%%%%%%%%%%%%%%%%%%%%%%%%%%%%%%%%%%%%

\section{Data acquisition or timing misbehavior}  

Since we are dealing with data acquisition or timing misbehavior,
it holds that $z_i(\cdot) \equiv x_i(\cdot)$
for every $i \in \I$. We will therefore only use $x_i$ throughout this section.
In order to prove Theorem 2, we avail ourselves of the following intermediate result. 
% Notice that for this result Assumption \ref{ass:connect_2} plays no role.

\begin{lemma} \label{lem:tec_3}
Consider the network system (\ref{eq:sys})-(\ref{eq:clock}),
with the misbehaving nodes exhibiting an error in the operation (i) and/or (iv). 
Suppose that Assumptions \ref{ass:M} and \ref{ass:solutions} hold. Let $T'$
be as in Lemma \ref{lem:tec_2}, and let $r$ and $s$ be
any two indices belonging to $\N$ such that 
$x_r(t)=\underline x$ and $x_s(t)=\overline x$ for all $t \geq T'$.
Then, $|\ave_r(t)| < 3 \eps/2$ and $|\ave_s(t)| < 3 \eps/2$ for all $t \geq T$,
where $T:=T' + \eps/4$.
\end{lemma}

\emph{Proof}.
We prove the claim only for node $r$ since the analysis for node $s$ is analogous.
Consider any sampling interval $T^r_k$ with $t^r_k \geq T'$.
As a first step, notice that $\ave_r(t) \geq 0$
for all $t \in T^r_k$ since, in view of Lemma \ref{lem:tec_2}, 
$j \in \D_r(t)$ only if $x_j(t) \geq \underline x$.
We stress that $\D_r(\cdot)$ is defined only for analysis purposes as its 
computation is done only at the update times.
We now determine an upper bound for $\ave_r(\cdot)$ over $T^r_k$.

Following the same notation as in Section 2.1, 
let $\underline {\mathcal E}_i(t)$ and $\overline {\mathcal E}_i(t)$
be the subset of nodes not belonging to $\D_i(t)$.
Decompose $\D_r(t) = \A_r(t) \cup \B_r(t) \cup \C_r(t)$, where
{\setlength\arraycolsep{1pt}
\begin{eqnarray} \label{lem:tec_3:eq_3}
\left\{
\begin{array}{l}
\A_r(t) := \D_r(t) \cap \underline {\mathcal E}_r(t^r_k) \\ 
\B_r(t) := \D_r(t) \cap {\D}_r(t^r_k) \\ 
\C_r(t) := \D_r(t) \cap \overline {\mathcal E}_r(t^r_k)
\end{array} 
\right.
\end{eqnarray}}%
Note that this can be done since $i \in  \D_r(t)$ only if $i \in \Q_r$
and since $\Q_r = \underline {\mathcal E}_r(t^r_k) \cup 
{\D}_r(t^r_k) \cup \overline {\mathcal E}_r(t^r_k)$
by construction.  {\color{black} The set $\C_r(t)$ is comprised of the neighborhood of node $r$ that had the highest values at time $t_k^r$, but have moderate values at time $t$.}
Further decompose $\C_r(t) = \underline \C_r(t) \cup \overline \C_r(t)$, where
{\setlength\arraycolsep{1pt}
\begin{eqnarray} \label{lem:tec_3:eq_5}
\left\{
\begin{array}{l}
\underline \C_r(t) := \{ j \in \C_r(t): x_j(t) = \underline x \} \\ 
\overline \C_r(t) := \{ j \in \C_r(t): x_j(t) > \underline x \}
\end{array} 
\right.
\end{eqnarray}}%
%the set C_r(t) is the set of nodes that had the highest values in r’s neighborhood at time t_k^r, but have moderate values at time t.  Based on this description, it becomes much easier to see that if \bar{C}_r has L nodes, then there must be at least L nodes whose values are more extreme at time t, and those nodes must have come from the set of moderate (or low) values from time t_k^r.
This can be done since $j \in \C_r(t)$ only if $j \in \D_r(t)$ and
$j \in \D_r(t)$ only if $x_j(t) \geq \underline x$.
We focus on the set $\overline \C_r(t)$. 
Suppose that there are $L$ elements in this set. 
Obviously $L \leq F$ since $|\overline {\mathcal E}_r(\tau)| \leq F$ for all $\tau \in \R_{\geq 0}$.
Now, to each element of $\overline \C_r(t)$
there corresponds at least an element belonging
to $\mathcal Z_r(t) := \overline {\mathcal E}_r(t) \cap (\underline {\mathcal E}_r(t^r_k) \cup \D_r(t^r_k))$,
that is $|\mathcal Z_r(t)| \geq L$. {\color{black} In words, if $\bar{\C}_r(t)$ has $L$ nodes, then there must be at least $L$ nodes whose values are more extreme at time $t$, and those nodes must have come from the set of moderate (or low) values at time $t_k^r$}. In fact, 
$|\overline {\mathcal E}_r(t)| \leq |\overline {\mathcal E}_r(t^r_k)| - |\overline \C_r(t)| + |\mathcal Z_r(t)|$
by construction. 
Hence, if $|\mathcal Z_r(t)|<L$
one would have $|\overline {\mathcal E}_r(t)| < |\overline {\mathcal E}_r(t^r_k)| \leq F$
along with elements in $\D_r(t)$, those belonging to the set $\overline \C_r(t)$,
which take on a value larger than $\underline x$. However,
this is not possible in view of the control logic.
Since any element in $\overline \C_r(t)$
must take on a value not larger than the value taken on by any element in $\mathcal Z_r(t)$, 
we conclude that
{\setlength\arraycolsep{1pt}
\begin{eqnarray} \label{lem:tec_3:eq_7}
\sum_{j \in \C_r(t)} \left( x_j(t)- \underline x \right) 
\leq \sum_{j \in \mathcal Z_r(t)} \left( x_j(t)- \underline x \right)
\end{eqnarray}}%

As a final step, let 
$\mathcal Z_r(t) = \underline {\mathcal Z}_r(t) \cup \overline {\mathcal Z}_r(t)$, where
{\setlength\arraycolsep{1pt}
\begin{eqnarray} \label{lem:tec_3:eq_9}
\left\{
\begin{array}{l}
\underline {\mathcal Z}_r(t) := \overline {\mathcal E}_r(t) \cap \underline {\mathcal E}_r(t^r_k) \\
\overline {\mathcal Z}_r(t) :=  \overline {\mathcal E}_r(t)  \cap  \D_r(t^r_k) 
\end{array} 
\right.
\end{eqnarray}}%
Then, 
{\setlength\arraycolsep{1pt}
\begin{eqnarray} \label{lem:tec_3:eq_10}
&& \sum_{j \in \D_r(t)} \left( x_j(t)- \underline x \right)  \leq
\nonumber \\ 
&& \sum_{j \in (\A_r(t) \cup \underline {\mathcal Z}_r(t)) } \left( x_j(t)- \underline x \right) + 
\sum_{j \in (\B_r(t) \cup \overline {\mathcal Z}_r(t)) } \left( x_j(t)- \underline x \right) \nonumber \\ 
\end{eqnarray}}%
The first sum on the right side of (\ref{lem:tec_3:eq_10}) yields
{\setlength\arraycolsep{1pt}
\begin{eqnarray} \label{lem:tec_3:eq_11}
&& \sum_{j \in (\A_r(t) \cup \underline {\mathcal Z}_r(t)) } \left( x_j(t)- \underline x \right)  \nonumber \\ 
&& \qquad \quad \leq \sum_{j \in (\A_r(t) \cup \underline {\mathcal Z}_r(t)) } \left( x_j(t^r_k) - \underline x + t-t^r_k  \right)   \nonumber \\ 
&& \qquad \quad \leq | \underline {\mathcal E}_r(t^r_k) | ( t-t^r_k ) 
\end{eqnarray}}%
The first inequality follows since all the nodes, including the misbehaving ones,
take controls in $\{-1,0,1\}$. 
The second inequality follows since 
$(\A_r(t) \cup \underline {\mathcal Z}_r(t)) \subseteq  \underline {\mathcal E}_r(t^r_k)$
and because $j \in \underline {\mathcal E}_r(t^r_k)$ only if 
$x_j(t^r_k) < \underline x$.
The second sum on the right side of (\ref{lem:tec_3:eq_10}) yields
{\setlength\arraycolsep{1pt}
\begin{eqnarray} \label{lem:tec_3:eq_12}
&& \sum_{j \in (\B_r(t) \cup \overline {\mathcal Z}_r(t)) } \left( x_j(t)- \underline x \right)  \nonumber \\ 
&& \qquad \quad \leq \sum_{j \in (\B_r(t) \cup \overline {\mathcal Z}_r(t)) } \left( x_j(t^r_k) - \underline x + t-t^r_k  \right)   \nonumber \\ 
&& \qquad \quad \leq \eps + | \D_r(t^r_k) | ( t-t^r_k ) 
\end{eqnarray}}%
The last inequality follows since $(\B_r(t) \cup \overline {\mathcal Z}_r(t)) \subseteq  \D_r(t^r_k)$
and since $\sum_{j \in \S} \left( x_j(t^r_k)- \underline x \right) < \eps$ 
for every $\S \subseteq \D_r(t^r_k)$. 
In fact, 
$\sum_{j \in \D_r(t^r_k)} \left( x_j(t^r_k) - \underline x \right) < \eps$ 
since $r$ stays constant from $T'$ on.
Thus, in order for
$\sum_{j \in \S} \left( x_j(t^r_k) - \underline x \right) \geq\eps$ 
there should exist at least one node $j \in \D_r(t^r_k) \backslash \S$
such that $x_j(t^r_k)- \underline x < 0$. However, 
since $r$ is the node that attains the minimum value 
among the normal nodes then $j \in \D_r(t^r_k)$ only if $x_j(t^r_k) \geq \underline x$
otherwise it is discarded in view of Assumption \ref{ass:M}
and by construction of the control logic. 

Overall, we get
{\setlength\arraycolsep{1pt}
\begin{eqnarray} \label{lem:tec_3:eq_13}
&& \sum_{j \in \D_r(t)} \left( x_j(t)- \underline x \right)  \leq
\nonumber \\ 
&& \qquad \eps + ( | \underline {\mathcal E}_r(t^r_k) | + | \D_r(t^r_k) | ) ( t-t^r_k ) < \frac{3}{2} \eps
\end{eqnarray}}%
for all $t \in T^r_k$ since $| \underline {\mathcal E}_r(t^r_k) | + | \D_r(t^r_k) | \leq d_r$ and because
$t - t^r_k \leq \eps/(4d_r)$ for all $t \in T^r_k$. Finally, since the interval $T^r_k$ is generic,
we conclude that $|\ave_r(\cdot)|<3 \eps/2$ starting from the first update 
$t^r_k \geq T'$ of node $r$.
Since this occurs not later than $T' + \eps/(4d_r)$,
it holds that $|\ave_r(t)|<3 \eps/2$ for all $t \geq T$. 
\qedp

We note that Lemma \ref{lem:tec_3} strongly relies
on the fact that there is no control or transmission misbehavior.
In fact, in either case, neither (\ref{lem:tec_3:eq_11}) nor  (\ref{lem:tec_3:eq_12})
are valid.

We finally proceed with the proof of Theorem 2.

\emph{Proof of Theorem 2}. Let $T$ be as in Lemma \ref{lem:tec_3}, 
and denote by $r$ and $s$ any two indices belonging to $\N$ such that $x_r(t) = \underline x$ 
and $x_s(t) = \overline x$ for all $t \geq T$. 

Consider any $t \geq T$. We have
{\setlength\arraycolsep{2pt}
\begin{eqnarray} \label{thm_2:eq_2}
\ave_s(t) &=& \sum_{j \in \D_s(t)} \left( x_j(t)- \overline x \right) \nonumber \\
&=& \sum_{j \in \D_s(t)} \left( x_j(t)- \underline x \right) + |\D_s(t)| \left( \underline x - \overline x \right)
\end{eqnarray}}%
The sum term satisfies 
{\setlength\arraycolsep{2pt}
\begin{eqnarray} \label{thm_2:eq_3}
&& \sum_{j \in \D_s(t)} \left( x_j(t)- \underline x \right) \nonumber \\
&& = \sum_{j \in (\D_s(t) \backslash \D_r(t))} \left( x_j(t)- \underline x \right) + 
\sum_{j \in (\D_s(t) \cap \D_r(t))} \left( x_j(t)- \underline x \right)  \nonumber \\
&& < \sum_{j \in (\D_s(t) \backslash \D_r(t))} \left( x_j(t)- \underline x \right) + \frac{3}{2}\eps  
\end{eqnarray}}%
The inequality comes from the fact that $\ave_r(t) < 3 \eps/2$ 
in view of Lemma \ref{lem:tec_3}, and since 
$\sum_{j \in \S} \left( x_j(t)- \underline x \right) < 3 \eps/2$
for every $\S \subseteq \D_r(t)$
(\emph{cf.} the proof of Lemma \ref{lem:tec_3}). Hence, 
{\setlength\arraycolsep{2pt}
\begin{eqnarray} \label{thm_2:eq_4}
&& \ave_s(t) < \nonumber \\
&& \sum_{j \in (\D_s(t) \backslash \D_r(t))} \left( x_j(t)- \underline x \right) 
+ |\D_s(t)| \left( \underline x - \overline x \right) + \frac{3}{2}\eps  
\end{eqnarray}}%
Since $s$ is the node attaining the maximum value 
among the normal nodes, $j \in \D_s(t)$ only if $x_j(t) \leq \overline x$
otherwise it is discarded in view of Assumption \ref{ass:M}
and by construction of the control logic. Thus, $x_j(t) \leq  \overline x$ for all 
$t \geq T$ and all $j \in \D_s(t)$.
Hence,
{\setlength\arraycolsep{2pt}
\begin{eqnarray} \label{thm_2:eq_5}
\ave_s(t) < \left(  |\D_s(t) \backslash \D_r(t)| - |\D_s(t)| \right) 
(\overline x - \underline x) + \frac{3}{2}\eps
\end{eqnarray}}%

As a final step, notice that
{\setlength\arraycolsep{2pt}
\begin{eqnarray} \label{thm_2:eq_6}
|\D_s(t)| - |\D_s(t) \backslash \D_r(t)| =  |\D_s(t) \cap \D_r(t)| 
\end{eqnarray}}%
Following the same notation as in Section 2.1, 
let $\underline {\mathcal E}_i(t)$ and $\overline {\mathcal E}_i(t)$
be the set of nodes discarded by $i \in \N$ at time $t$ 
with associated value smaller than $x_i(t)$ and larger than $x_i(t)$, respectively.
For nodes $r$ and $s$, define
{\setlength\arraycolsep{1pt}
\begin{eqnarray} \label{thm_2:eq_7}
\left\{
\begin{array}{l}
\underline \W_r(t) :=  \underline {\mathcal E}_r(t) \cap \Q_s \\ 
\overline \W_r(t) :=  \overline {\mathcal E}_r(t) \cap \Q_s \\ 
\underline \W_s(t) :=  \underline {\mathcal E}_s(t) \cap \Q_r \\
\overline \W_s(t) :=  \overline {\mathcal E}_s(t) \cap \Q_r
\end{array} 
\right.
\end{eqnarray}}%
Thus, at every $t \in \R_{\geq 0}$, node $r$ discards 
$|\underline \W_r(t)| + |\overline \W_r(t)|$ nodes that are also
neighbors of $s$. Similarly, node $s$ discards 
$|\underline \W_s(t)| + |\overline \W_s(t)|$ nodes that are also
neighbors of $r$. Moreover, 
{\setlength\arraycolsep{1pt}
\begin{eqnarray} \label{thm_2:eq_8}
\left\{
\begin{array}{l}
\underline \W_r(t) \subseteq \underline \W_s(t) \\ 
\overline \W_s(t) \subseteq \overline \W_r(t) 
\end{array} 
\right.
\end{eqnarray}}%
The first relation follows because $r$ is the node attaining the
minimum value among the normal nodes. Thus, all the nodes that belong to 
$\underline \W_r(t)$ take on value less than $\underline x$ and hence are necessarily misbehaving. 
Since these nodes belong to $\Q_s$, they must be discarded also by node $s$.
In fact, at every $t \in \R_{\geq 0}$,
node $s$ discards the $F$ smallest value less than $x_s(t)$,
and, after $T$, there cannot be more than $F$ values less than $\underline x$ in view of 
Assumption \ref{ass:M} and Lemma \ref{lem:tec_2}.
Hence, these nodes must belong to $\underline {\mathcal E}_s(t)$, and thus to $\underline \W_s(t)$.
The same reasoning applies to the relation $\overline \W_s(t) \subseteq \overline \W_r(t)$.

Hence, at every $t \in \R_{\geq 0}$, nodes $r$ and $s$ can discard at most
$|\underline \W_s(t)| + |\overline \W_r(t)| \leq 2F$ different common neighbors.
Since by Assumption \ref{ass:connect_2} nodes $r$ and $s$ have at least 
$2F+1$ neighbors in common, we have $|\D_s(t) \cap \D_r(t)| \geq 1$.
This implies that $\ave_s(t) < - (\overline x - \underline x) + 3 \eps/2$. 
By combining this inequality with $\ave_s(t) > -3 \eps/2$, we finally conclude that 
$\overline x - \underline x < 3 \eps$. \qedp
 
%%%%%%%%%%%%%%%%%%%%%%%%%%%%%%%%%%%%%%%%%%%%%%%%%%%%%%%%%%%%%%%%%%%%%%%%%%%%%%%%
%%%%%%%%%%%%%%%%%%%%%%%%%%%%%%%%%%%%%%%%%%%%%%%%%%%%%%%%%%%%%%%%%%%%%%%%%%%%%%%%

%%%%%%%%%%%%%%%%%%%%%%%%%%%%%%%%%%%%%%%%%%%%%%%%%%%%%%%%%%%%%%%%%%%%%%%%%%%%%%%%
%%%%%%%%%%%%%%%%%%%%%%%%%%%%%%%%%%%%%%%%%%%%%%%%%%%%%%%%%%%%%%%%%%%%%%%%%%%%%%%%

\section{A numerical example}
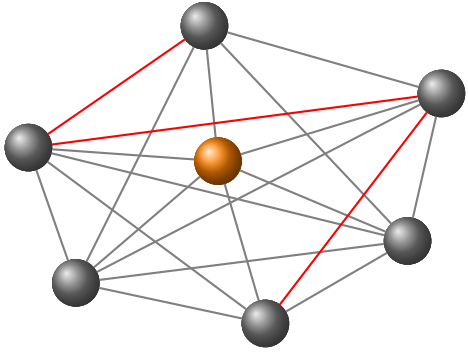
\begin{figure}
\begin{center}
\begin{tikzpicture}[scale=0.9]
\GraphInit[vstyle=Art]
\tikzset{VertexStyle/.style = {shape = circle, ball color = gray,
text = black, inner sep = 2pt, outer sep = 0pt, minimum size = 18 pt}}
\tikzset{EdgeStyle/.style = {-, gray}   }
    \Vertex[x=-1.1,y=1.2]{A}
	\Vertex[x=2.4, y=-1.4]{B}
	\Vertex[x=5,y=2]{D}
	\Vertex[x=1.5, y=3]{E}
	\tikzset{VertexStyle/.style = {shape = circle, ball color =orange,
text = black, inner sep = 2pt, outer sep = 0pt, minimum size = 18 pt}}
    \Vertex[x=1.7, y=1.]{G}
		\tikzset{VertexStyle/.style = {shape = circle, ball color =gray,
text = black, inner sep = 2pt, outer sep = 0pt, minimum size = 18 pt}}
	\Vertex[x=-0.4, y=-0.8]{F}
    \Vertex[x=4.5, y=-.18]{H}
\Edge[](A)(F) \Edge[](A)(G) \Edge[](A)(H)  
\Edge[](B)(A) \Edge[](B)(F) \Edge[](B)(G) \Edge[](B)(H)
\Edge[](D)(E)	\Edge[](D)(F) \Edge[](D)(G) \Edge[](D)(H)
\Edge[](E)(F) \Edge[](E)(G) \Edge[](E)(H)
\Edge[](F)(G) \Edge[](F)(H) \Edge[](G)(H)
\tikzset{EdgeStyle/.style = {-, red}   }
\Edge[](B)(D) \Edge[](A)(D) \Edge[](A)(E)

%\GraphInit[vstyle=Art]
%\tikzset{VertexStyle/.style = {shape = circle, ball color = orange,
%text = black, inner sep = 2pt, outer sep = 0pt, minimum size = 7 pt}}
  %\Vertex[x=-0, y=2.8]{A}
	%\Vertex[x=1.5, y=1.6]{B}
	%\Vertex[x=0.4, y=0.6]{C}
	%\Vertex[x=-0.8, y=0.3]{D}
	%\Vertex[x=-0.8, y=1.6]{E}
	%\tikzset{VertexStyle/.style = {shape = circle, ball color = gray,
%text = black, inner sep = 2pt, outer sep = 0pt, minimum size = 7 pt}}
	%\Vertex[x=1.5, y=3.3]{F}
	%\Vertex[x=2.4, y=0]{G}	
 	%\Vertex[x=0.1, y=-1.1]{H}
 	%\Vertex[x=-2.3, y=0.8]{I}	
	%\Vertex[x=-1.5, y=3.3]{J}
			%%\tikzset{VertexStyle/.style = {shape = circle, ball color =orange,
%%text = black, inner sep = 2pt, outer sep = 0pt, minimum size = 9 pt}}
    %%\Vertex[x=1.7, y=1.]{G}
		%\tikzset{EdgeStyle/.style = {-, gray}}
		 %Compltet graph
		%\Edge[](A)(B) \Edge[](A)(C) \Edge[](A)(D) \Edge[](A)(E)
		%\Edge[](B)(C) \Edge[](B)(D) \Edge[](B)(E) 
		%\Edge[](C)(D) \Edge[](C)(E) 
		%\Edge[](D)(E)
		 %The rest of the graph
   %\Edge[](F)(A) \Edge[](F)(B) \Edge[](F)(C) \Edge[](F)(D) \Edge[](F)(E)
	 %\Edge[](G)(A) \Edge[](G)(B) \Edge[](G)(C) \Edge[](G)(D) \Edge[](G)(E)
	 %\Edge[](H)(A) \Edge[](H)(B) \Edge[](H)(C) \Edge[](H)(D) \Edge[](H)(E)
	 %\Edge[](I)(A) \Edge[](I)(B) \Edge[](I)(C) \Edge[](I)(D) \Edge[](I)(E)
	 %\Edge[](J)(A) \Edge[](J)(B) \Edge[](J)(C) \Edge[](J)(D) \Edge[](J)(E)
%\Edge[](B)(A) \Edge[](B)(F) \Edge[](B)(G) \Edge[](B)(H)
%\Edge[](D)(E)	\Edge[](D)(F) \Edge[](D)(G) \Edge[](D)(H)
%\Edge[](E)(F) \Edge[](E)(G) \Edge[](E)(H)
%\Edge[](F)(G) \Edge[](F)(H) \Edge[](G)(H)
%\tikzset{EdgeStyle/.style = {-, red}   }
%\Edge[](B)(D) \Edge[](A)(D) \Edge[](A)(E)

\end{tikzpicture}
\end{center}
\caption{Network system considered in the numerical example.
Normal nodes are depicted in grey, while the misbehaving node is depicted in orange.
The graph satisfies Assumption \ref{ass:connect_1} and is thus robust against 
generic misbehavior (Theorem \ref{thm_1}). The removal of the red edges leads to a graph that 
satisfies Assumption \ref{ass:connect_2}, which is robust against 
data acquisition or timing misbehavior (Theorem \ref{thm_2}). }
\label{fig:graph}
\end{figure}

Consider a network system as in (\ref{eq:sys})-(\ref{eq:clock}), with $n=7$ nodes interconnected 
as in Fig. \ref{fig:graph} and $F=1$ misbehaving nodes. 
State and clock initial values are taken randomly 
within the intervals $[0,1]$ and $[0, t_{init}]$, respectively, 
with $t_{init}=0.15 sec$. The desired accuracy level for consensus is selected as $\eps=0.01$ and we set $\underline \Delta_i=\eps /(4 d_i)$ for every node. 
We note that the graph satisfies Assumption \ref{ass:connect_1}, which 
is sufficient to guarantee resilience against generic misbehavior (Theorem \ref{thm_1}). 

\begin{figure}
	\centering
		\includegraphics[width=0.5\textwidth]{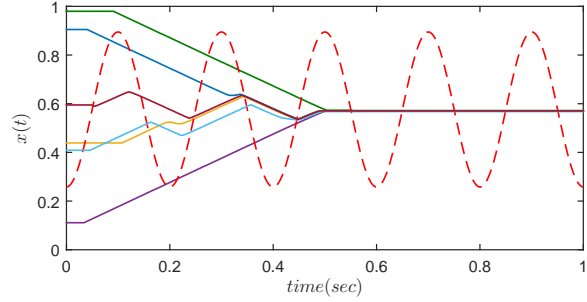}
	\caption{Network state evolution with the resilient consensus protocol 
	under control misbehavior. The evolution of the 
	misbehaving node is depicted in red dashed line.}
	\label{fig:fig2}
\end{figure}

We consider the case of control misbehavior, which is most critical for consensus
(\emph{cf.} Section \ref{sec:ass}). Specifically, we assume that the misbehaving 
node applies the control input $u_i(t)=10\sin(10 \, \pi t)$ for all $t \in \R_{\geq0}$ instead of (\ref{eq:controls}). 
Fig. \ref{fig:fig2} illustrates the network state evolution with the proposed resilient 
consensus protocol. In agreement with the conclusions of Theorem \ref{thm_1}, one 
sees that the normal nodes remain in the convex hull containing their initial values
and reach a practical agreement disregarding the behavior of the misbehaving node. 

%%%%%%%%%%%%%%%%%%%%%%%%%%%%%%%%%%%%%%%%%%%%%%%%%%%%%%%%%%%%%%%%%%%%%%%%%%%%%%%%
%%%%%%%%%%%%%%%%%%%%%%%%%%%%%%%%%%%%%%%%%%%%%%%%%%%%%%%%%%%%t%%%%%%%%%%%%%%%%%%%%

\section{Conclusions}

This paper shows the possibility of approaching the resilient consensus 
problem in a context where the nodes have their own clocks
and can make updates at arbitrary time instants. Although the
results are preliminary, they indicate that handling misbehaving units
can be possible also in network applications involving 
asynchronous and aperiodic transmissions, as occurs with 
event-triggered and self-triggered network systems. 

We have considered a scenario where the network can support the
data flow with {reliability} and {accuracy}, thus
neglecting issues such as transmission delays,
data loss, bandwidth as well as noise. In practice,
these issues are are also very important but require careful 
consideration of several technicalities. Nonetheless,
we envision that some extensions are indeed possible along the same
lines as in \cite{de2013robust,senejohnny2017jamming} where 
we discuss aspects related to the quality of the transmission medium.

Our approach utilizes a self-triggered update scheme with
control saturation. It is of interest to investigate if similar results can be obtained
also with event-triggered or other types of aperiodic update schemes \cite{HetelAUT2017}.
It is also interesting to see if similar results can be obtained with 
other averaging functions, for example with the classic coupling law
for \emph{average} consensus \cite{cortes2006finite}.
We envision an application of the present research within the context of
distributed optimization \cite{SundaramAllerton2015}, with 
specific reference to self-triggered schemes \cite{FazlyabCDC2016}.
Another interesting research venue is in the area of
multi-agent systems with cloud access \cite{NowzariACC2016}. 
Also in this context, self-triggered control seems a viable option
for enabling asynchronous coordination without destroying regulation properties.

\bibliographystyle{IEEETran} 

\bibliography{Automatica_Draft}

\begin{thebibliography}{10}
\providecommand{\url}[1]{#1}
\csname url@samestyle\endcsname
\providecommand{\newblock}{\relax}
\providecommand{\bibinfo}[2]{#2}
\providecommand{\BIBentrySTDinterwordspacing}{\spaceskip=0pt\relax}
\providecommand{\BIBentryALTinterwordstretchfactor}{4}
\providecommand{\BIBentryALTinterwordspacing}{\spaceskip=\fontdimen2\font plus
\BIBentryALTinterwordstretchfactor\fontdimen3\font minus
  \fontdimen4\font\relax}
\providecommand{\BIBforeignlanguage}[2]{{%
\expandafter\ifx\csname l@#1\endcsname\relax
\typeout{** WARNING: IEEEtran.bst: No hyphenation pattern has been}%
\typeout{** loaded for the language `#1'. Using the pattern for}%
\typeout{** the default language instead.}%
\else
\language=\csname l@#1\endcsname
\fi
#2}}
\providecommand{\BIBdecl}{\relax}
\BIBdecl

\bibitem{cao2013overview}
Y.~Cao, W.~Yu, W.~Ren, and G.~Chen, ``An overview of recent progress in the
  study of distributed multi-agent coordination,'' \emph{IEEE Transactions on
  Industrial informatics}, vol.~9, no.~1, pp. 427--438, 2013.

\bibitem{Dolev1986}
D.~Dolev, N.~A. Lynch, S.~S. Pinter, E.~W. Stark, and W.~E. Weihl, ``Reaching
  approximate agreement in the presence of faults,'' \emph{Journal of the ACM},
  vol.~33, no.~3, pp. 499--516, 1986.

\bibitem{Lynch}
N.~A. Lynch, \emph{Distributed Algorithms}.\hskip 1em plus 0.5em minus
  0.4em\relax Morgan Kaufmann, 1996.

\bibitem{Sundaram2013}
H.~J. LeBlanc, H.~Zhang, X.~Koutsoukos, and S.~Sundaram, ``Resilient asymptotic
  consensus in robust networks,'' \emph{IEEE Journal on Selected Areas in
  Communications,}, vol.~31, no.~4, pp. 766--781, 2013.

\bibitem{SaldanaACC2017}
D.~{Salda\~na}, A.~Prorok, S.~Sundaram, M.~F.~M. Campos, and V.~Kumar,
  ``Resilient consensus for time-varying networks of dynamic agents,'' in
  \emph{2017 American Control Conference}.\hskip 1em plus 0.5em minus
  0.4em\relax IEEE, 2017, pp. 2378--5861.

\bibitem{DibajiSCL2015}
S.~M. Dibaji and H.~Ishii, ``Consensus of second-order multi-agent systems in
  the presence of locally bounded faults,'' \emph{Systems \& Control Letters},
  vol.~79, pp. 23--29, 2015.

\bibitem{Abbas2014}
W.~Abbas, Y.~Vorobeychik, and X.~Koutsoukos, ``Resilient consensus protocol in
  the presence of trusted nodes,'' in \emph{International Symposium on
  Resilient Control Systems}.\hskip 1em plus 0.5em minus 0.4em\relax IEEE,
  2014.

\bibitem{Usevitch2017}
J.~Usevitch and D.~Panagou, ``r-robustness and (r,s)-robustness of circulant
  graphs,'' in \emph{arxiv.org/abs/1710.01990}, 2017.

\bibitem{LeBlanc2012}
H.~J. LeBlanc and X.~Koutsoukos, ``Resilient asymptotic consensus in
  asynchronous robust networks,'' in \emph{50th Annual Allerton Conference on
  Communication, Control, and Computing}.\hskip 1em plus 0.5em minus
  0.4em\relax IEEE, 2012, pp. 1742--1749.

\bibitem{Vaidya2012}
N.~H. Vaidya, L.~Tseng, and G.~Liang, ``Iterative approximate byzantine
  consensus in arbitrary directed graphs,'' in \emph{ACM Symposium on
  Principles of Distributed Computing}.\hskip 1em plus 0.5em minus 0.4em\relax
  IEEE, 2012, pp. 365--374.

\bibitem{DibajiAUT2017}
S.~M. Dibaji and H.~Ishii, ``Resilient consensus of second-order agent
  networks: Asynchronous update rules with delays,'' \emph{Automatica},
  vol.~81, pp. 123--132, 2017.

\bibitem{DibajiTAC2017}
S.~M. Dibaji, H.~Ishii, and R.~Tempo, ``Resilient randomized quantized
  consensus,'' \emph{IEEE Transactions on Automatic Control}, vol.~PP, pp. 1--1
  (In press), 2017.

\bibitem{heemels2012introduction}
W.~Heemels, K.~H. Johansson, and P.~Tabuada, ``An introduction to
  event-triggered and self-triggered control,'' in \emph{Decision and Control
  (CDC), 2012 IEEE 51st Annual Conference on}.\hskip 1em plus 0.5em minus
  0.4em\relax IEEE, 2012, pp. 3270--3285.

\bibitem{de2013robust}
C.~De~Persis and P.~Frasca, ``Robust self-triggered coordination with ternary
  controllers,'' \emph{IEEE Transactions on Automatic Control}, vol.~58,
  no.~12, pp. 3024--3038, 2013.

\bibitem{cortes2006finite}
J.~Cort{\'e}s, ``Finite-time convergent gradient flows with applications to
  network consensus,'' \emph{Automatica}, vol.~42, no.~11, pp. 1993--2000,
  2006.

\bibitem{FranceschelliTAC2017}
M.~Franceschelli, A.~Giua, and A.~Pisano, ``Finite-time consensus on the median
  value with robustness properties,'' \emph{IEEE Transactions on Automatic
  Control}, vol.~62, no.~4, pp. 1652--1667, 2017.

\bibitem{Godsil}
C.~Godsil and G.~Royle, \emph{Algebraic Graph Theory}.\hskip 1em plus 0.5em
  minus 0.4em\relax Springer-Verlag New York, 2001.

\bibitem{senejohnny2017jamming}
D.~M. Senejohnny, P.~Tesi, and C.~{De Persis}, ``A jamming-resilient algorithm
  for self-triggered network coordination,'' \emph{IEEE Transactions on Control
  of Network Systems}, vol.~PP, pp. 1--1 (In press), 2017.

\bibitem{HetelAUT2017}
L.~Hetel, C.~Fiter, H.~Omran, A.~Seuret, E.~Fridman, J.~Richard, and S.~I.
  Niculescu, ``Recent developments on the stability of systems with aperiodic
  sampling: An overview,'' \emph{Automatica}, vol.~76, pp. 309--3350, 2017.

\bibitem{SundaramAllerton2015}
S.~Sundaram and B.~Gharesifard, ``Consensus-based distributed optimization with
  malicious nodes,'' in \emph{53rd Annual Allerton Conference on Communication,
  Control, and Computing}.\hskip 1em plus 0.5em minus 0.4em\relax IEEE, 2015,
  pp. 244--249.

\bibitem{FazlyabCDC2016}
M.~Fazlyab, C.~Nowzari, G.~J. Pappas, A.~Ribeiro, and V.~M. Preciado,
  ``Self-triggered time-varying convex optimization,'' in \emph{55th Conference
  on Decision and Control}.\hskip 1em plus 0.5em minus 0.4em\relax IEEE, 2016,
  pp. 3090--3097.

\bibitem{NowzariACC2016}
C.~Nowzari and G.~J. Pappas, ``Multi-agent coordination with asynchronous cloud
  access,'' in \emph{2016 American Control Conference}.\hskip 1em plus 0.5em
  minus 0.4em\relax IEEE, 2016, pp. 4649--4654.

\end{thebibliography}

\end{document}